\documentclass[sigconf]{acmart}

\copyrightyear{2025}
\acmYear{2025}
\setcopyright{cc}
\setcctype{by}
\acmConference[CHI '25]{CHI Conference on Human Factors in Computing Systems}{April 26-May 1, 2025}{Yokohama, Japan}
\acmBooktitle{CHI Conference on Human Factors in Computing Systems (CHI '25), April 26-May 1, 2025, Yokohama, Japan}\acmDOI{10.1145/3706598.3713256}
\acmISBN{979-8-4007-1394-1/25/04}

\begin{document}

\title[``Impressively Scary:'' User Perceptions and Reactions to AI/ML Models Within Social Media Applications]{``Impressively Scary:'' Exploring User Perceptions and Reactions to Unraveling Machine Learning Models in Social Media Applications}

\author{Jack West}
\affiliation{%
  \institution{University of Wisconsin-Madison}
  \city{Madison WI}
  \country{USA}
}
\email{jwwest@wisc.edu}

\author{Bengisu Cagiltay}
\affiliation{%
  \institution{University of Wisconsin-Madison}
  \city{Madison WI}
  \country{USA}
}
\email{bengisu@cs.wisc.edu}

\author{Shirley Zhang}
\affiliation{%
  \institution{University of Wisconsin-Madison}
  \city{Madison WI}
  \country{USA}
}
\email{hzhang664@wisc.edu}

\author{Jingjie Li}
\affiliation{%
  \institution{University of Edinburgh}
  \city{Edinburgh}
  \country{UK}
}
\email{jingjie.li@ed.ac.uk}

\author{Kassem Fawaz}
\affiliation{%
  \institution{University of Wisconsin-Madison}
  \city{Madison WI}
  \country{USA}
}
\email{kfawaz@wisc.edu}

\author{Suman Banerjee}
\affiliation{%
  \institution{University of Wisconsin-Madison}
  \city{Madison WI}
  \country{USA}
}
\email{suman@cs.wisc.edu}


\begin{abstract}
    Machine learning models deployed locally on social media applications are used for features, such as face filters which read faces in-real time, and they expose sensitive attributes to the apps.
    However, the deployment of machine learning models, e.g., when, where, and how they are used, in social media applications is opaque to users.
    We aim to address this inconsistency and investigate how social media user perceptions and behaviors change once exposed to these models.
    We conducted user studies (N=21) and found that participants were unaware to both \textit{what} the models output and \textit{when} the models were used in Instagram and TikTok, two major social media platforms.
    In response to being exposed to the models' functionality, we observed long term behavior changes in 8 participants.
    Our analysis uncovers the challenges and opportunities in providing transparency for machine learning models that interact with local user data.
\end{abstract}

\maketitle

\section{Introduction}

Social media apps have become integral to modern life, with 4.95 billion users connecting through mobile and web platforms~\cite{backlinkoSocialNetwork}. Billions of people use mobile devices to access social media, with Instagram and TikTok boasting over one billion~\cite{googleTikTokApps} and five billion~\cite{googleInstagramApps} downloads, respectively.
One of the key factors to their popularity is content tailored to individual preferences. As innovators in the technology space~\cite{facebookPublicationsMeta,tiktokTikToksPrivacy}, social media companies are leaders in recommendation algorithms~\cite{tiktokTikTokRecommends,instagramLoginx2022} and machine learning (ML) models~\cite{metaIntroducingMeta,metaDINOv2Stateoftheart}. To train these models, social media companies utilize user data~\cite{businessinsiderMetaUses}.

Artificial Intelligence/ Machine Learning (AI/ML) models in social media \textit{already} have a multi-billion dollar market size and are expected to continually grow~\cite{marketsandmarketsSocialMedia,mordorintelligenceSocialMedia,trammell2023economic}. AI/ML advances are happening at an unprecedented pace~\cite{yahooMovingFast,fastAIWhat}, limiting users' access to accurate and up-to-date information and impacting companies to understand how and when AI is employed~\cite{jones2022ai,ouchchy2020ai}. In an article written in Meta's Transparency Center~\cite{metaInstagramFeed}, they describe how their AI/ML models are used in their algorithm, \textit{``The AI system considers a variety of \textbf{input signals} about each post. These signals might include who created the post or the type of content in the post. A lightweight model is run to select approximately 500 of the most relevant posts.''} Transparency statements from companies seldom explain \textit{when} they apply these algorithms and \textit{what} specific input signals they take into account.

While social media companies may not be legally obligated to be fully transparent about their algorithms (e.g., due to intellectual property concerns), the lack of transparency propels users to rely on their perceptions to explain what they see on social media~\cite{rader2018explanations}. Documenting and understanding these algorithmic perceptions is a well-established field in the HCI community~\cite{myers2018censored,devito2018people,fletcher2019generalised,delmonaco2024you,devito2017algorithms,taylor2022initial,register2023attached,zhao2016social}.
Prior works about social media algorithm perceptions focus on advertisements~\cite{saha2021advertiming,sharma2023user}, social media feeds~\cite{eslami2015always,devito2017algorithms,devito2018people}, misinformation~\cite{chen2022visualbubble,wischnewski2021disagree,chen2023spread}, and several other social media features relying on algorithm decision making~\cite{fletcher2019generalised,delmonaco2024you,cotter2019playing}. Prior studies explore models based on patents~\cite{andalibi2020human} or focused on feed curation~\cite{epstein2022yourfeed}, and they have not explicitly exposed the functionality of authentic models to study participants. As a result, there is a gap in understanding the actual user reactions and perceptions towards \textit{authentic} models employed by social media.

Here, we leverage recent advances in reverse engineering of mobile apps to enable user interactions with authentic AI/ML models from Instagram and TikTok~\cite{west2024picture}. With access to these authentic models, we aim to explore user perceptions and behavior changes when exposed to transparent AI/ML functionality. In particular, we ask the following research questions:

\begin{itemize}
    \item \textbf{RQ1}: What are social media users' \textit{prior understandings and perceptions} toward the usage of ML/AI within social media applications?
    \item \textbf{RQ2}: What are users’ \textit{immediate reactions} to and understandings from our intervention, which reveals how ML/AI models use and analyze local data on Instagram and TikTok?
    \item \textbf{RQ3}: How do users’ understanding of ML and associated \textit{behaviors change} after learning about authentic models deployed by Instagram and TikTok?
\end{itemize}

To address these research questions, we conducted an exploratory user study with 21 social media users and evaluated how their perceptions and behaviors change once exposed to authentic AI/ML models. In our study, participants interacted with the models and explored \textit{what} and \textit{how} their data is being processed locally by social media companies.

First, we conducted an interview to understand the participants' \textit{prior} perceptions and behaviors. We then demonstrated a basic example of computer vision using Google's Teachable Machine~\cite{carney2020teachable} and discussed concepts of computer vision.
Second, we demonstrated two computer vision models (from Instagram and TikTok) and allowed the participant to interact directly with the model.
A novel contribution of our work is the use of authentic AI/ML models and classifying participant data in real-time.
To address the final research question, during the interview, we asked participants if they thought any behaviors may change after seeing the models.
We then distributed a follow-up survey two weeks after the last participant to see if any reported behavior changes persisted.

Our findings reveal a mismatch between participants' expectations and how Instagram and TikTok actually employ AI/ML. They expected social media platforms to employ AI/ML on data they intentionally provide to the apps, such as posts and likes. However, direct exposure to authentic vision models heightened concerns around data collection practices, particularly regarding demographic information. These concerns depended on how participants interacted with the app. Participants reacted negatively to TikTok's model analyzing their camera frames without notification, explaining that it felt \textit{non-consensual}. On the other hand, participants felt uncomfortable with the abundance of data Instagram extracted from their images (over 500+ different abstract concepts from a single image). They were confused about what Instagram's data would be useful for as the amount was \textit{overwhelming} and the data's relevance to the app was \textit{unclear}. While participants exhibited varying attitudes about the local processing of their data, they all expressed interest in increased transparency from social media apps. However, they had different preferences regarding what information the apps should make available, such as AI/ML inputs, outputs, purposes, timing, and controls. Further, our work explores user-reported \textit{short-term} and \textit{long-term} change. Interestingly, two weeks after the conclusion of our study, eight users (out of 17 who filled out our post-study survey) reported a change in their behaviors.

\section{Related Work}
Our work aims to fill knowledge gaps in several areas in the HCI community.
Specifically, we address gaps in AI/ML algorithms in social media, XAI and transparency, user perceptions towards algorithms in the context of social media, and user privacy perceptions.

\subsection{Understanding of AI/ML Algorithms}

A large body of work studies users' understandings and reactions to AI/ML~\cite{ashktorab2021effects,andalibi2020human,gu2024data,kapania2022because,shandilya2022understanding,zhang2021did,poursabzi2021manipulating,zehrung2021vis,li2024machine,yin2019understanding}.
The existing works suggest that participants have mixed reactions, which depend on context. 
For example, Ashktorab et al.~\cite{ashktorab2021effects} performed a large-scale user study where participants were paired with AI in a game where the parties could each have the role of question giver or answering the questions.
Participants' likeability towards AI significantly dropped when the participant needed to depend on the AI to guess questions correctly.
However, participants reacted more positively when the AI was the question giver.
Ashktorab showed that \textit{context matters} when understanding reactions to AI.
To further the idea that context impacts user reactions to AI, Kapania et al.~\cite{kapania2022because} interviewed citizens of India and found that people were overwhelmingly positive towards AI decisions, as they felt they would be more correct than human decisions.
However, users tend to have negative attitudes toward the usage of AI in algorithms~\cite{rae2024effects,andalibi2020human,Dietvorst_2015,zhang2021did}.
Zhang et al.~\cite{zhang2021did} interviewed several users about their attitudes toward computer vision monitoring.
They found that users were most likely to be somewhat uncomfortable (36\%) with video analysis done on them. 
We expand on these works by establishing what specific details about modern AI/ML users do not like about model deployment.
We show that AI/ML sentiments also apply to \textit{implementation method}, not just results as indicated by these works.

\subsubsection{Understanding AI/ML Algorithms in Social Media}
Exposing participants to underlying algorithms in social media has been extensively studied~\cite{jahanbakhsh2023exploring,eslami2015always,eslami2016first,andalibi2020human,feng2024mapping}.
Feng et al.~\cite{feng2024mapping} found that people \textit{did not trust} the algorithm to give them the best content possible.
They thought that an algorithm would misunderstand or miss simple nuances.
Jahanbakhsh et al.~\cite{jahanbakhsh2023exploring} found participants' agreement rate of AI curated content fell once they were told the posts were selected by an AI. 
Feng et al. and Jahanbakhsh et al. show that participants lack trust in AI personalization.
Andalibi et al.~\cite{andalibi2020human} demonstrated users' observed reactions to emotion recognition ML models in the context of social media.
They found that several users had overwhelmingly negative reactions toward having their emotions analyzed, as they felt it was too personal.
Our work builds upon these and differentiates itself in three ways: first, we present participants with real-world models of how AI/ML models are used for image classification, offering a more practical demonstration. 
Second, our approach is specifically tailored to social media platforms such as Instagram and TikTok, grounding our discussion in users' direct experiences.
Third, we present AI/ML models to users separate from any algorithms, focusing on how participants felt about it being used in a more general sense.

\subsubsection{Ecological Validity in AI/ML Research In Social Media}
While there exists work that utilizes reverse engineering techniques for security studies on smartphones~\cite{van2018x,balebako2013little} similar methods have yet to be applied to AI/ML in social media.
Research in social media has inherent trade-offs on ecological validity~\cite{salehzadeh2023changes,bandy2021more} due to the opaqueness of the algorithms they explore.
Because of this limited information, similar studies relied on patents~\cite{andalibi2020human,roemmich2021data} or simulations~\cite{saha2021advertiming,chen2022visualbubble} to reconstruct AI/ML functionality. As a result, prior works that studied image analysis of users~\cite{zhang2021did} lacked the prior experience and private reflections that our study portrays. 
Works that specifically deal with authentic algorithms~\cite{eslami2015always} also exhibited the same reflective thought process. Participants reacted more strongly to models operating \textit{within the camera} without their knowledge.
Authentic models within social media apps offer two unique nuances: personal (1) reflections and (2) reactions. %

\subsection{AI Explainability and Transparency}
AI explainability~\cite{kim2023help,weisz2024design,liao2021should,ehsan2023charting,yuan2023contextualizing,ribeiro2016should,yildirim2024sketching,feng2024mapping,ehsan2021expanding} (XAI) is a body of work that explores how AI methods should be explained to user of an application.
AI transparency~\cite{liao2021should,lee2018understanding,sivertsen2024machine,eslami2019user,el2024transparent,shin2022algorithm,felzmann2019transparency,shusas2024trust,arnold2019factsheets} discusses the impact of applying and methods for implementing transparency for AI.
Our work highlights aspects of both topics.
Liao et al. ~\cite{liao2021should} described how AI systems should inform users about how and why their personal information is collected. 
They found that more transparency could translate into a higher sense of user trust.
Further, AI transparency and explainability are only becoming more relevant with the HCI community as the European AI transparency laws~\cite{euaiactIssueTransparency} are implemented.
El et al.~\cite{el2024transparent} hosted a workshop discussing potential directions and important questions for the HCI community regarding both transparency and explainability.
While the work's context surrounds generative AI, we found that some of the broader questions can also be applied to other types of AI/ML models.
Ehsan et al.~\cite{ehsan2024xai} demonstrated that explainability is relative to the person using the tool.
They claim that those who are more comfortable with AI are able to interpret the tools better.
Our work contributes to the ongoing discussion around AI explainability and transparency by addressing user understanding in the context of social media. 
While prior studies highlight the importance of transparency in increasing user trust ~\cite{liao2021should}, explainability varies based on user's familiarity ~\cite{ehsan2024xai}, we extend the research area by exploring how the social media environment affects AI explanations.

\subsection{AI/ML Privacy} 
Recent studies have explored new privacy and trust concerns around quickly advancing AI/ML capabilities~\cite{chiang2024more,asthana2024know,lee2024deepfakes,vereschak2024trust,zhang2024sa}.
Vereschak et al.~\cite{vereschak2024trust} found that for AI/ML models to be trusted, the models must have positive expectations from the user.
Vereschak et al. also argued that this trust depends on the provider of the model.
Asthana et al. ~\cite{asthana2024know} explored the question of how participants felt about certain model inferences on their data.
They found that participants were least comfortable with data being collected on \textit{protected} attributes such as race.
Our work examines the privacy perceptions towards local machine learning models and defines new problems that stem from local AI/ML processing.

\section{Methods}

We conducted and analyzed semi-structured interviews with adult social media users (N=21) in the U.S.
We include our recruitment material, screening form, interview protocol, and post-interview survey in an OSF repository for open access. \footnote{\url{https://osf.io/m4nru/?view_only=29ec93cc559d46af94d22ea9b100a4df}}

{\footnotesize
\begin{table*}[t]
\centering
\caption{ \textcolor{black}{Participant  demographics and tech savviness (Self reported, rated on a 5-point scale)}}
\Description{Table 1 is a list of all participants. Each participant's age, identified gender, highest completed degree and self indicated tech savviness. The caption for the table reads: "A table consisting of all participants with relevant demographic information. Participants were asked to report how tech savvy they felt they were. We asked them on a scale from 1 to 5. The numbers in the tech savvy column are the numbers they reported.}
\begin{tabular}{@{}lllll@{}}
\toprule
Participant ID & Identified Gender & Age Bin &  Highest Education Completed & Tech Savviness\\ 
\midrule
P1         &  Male & 26-35 & PhD & 4 \\
P2         &  Female & 26-35 & Masters Degree & 3\\
P3         &  Female & 26-35 & Masters Degree & 3\\
P4         &  Male & 18-25 & Undergraduate Degree & 3 \\
P5         &  Female & 36-45 & Masters Degree & 4 \\
P6         &  Male & 36-45 & Masters Degree & 4 \\
P7         &  Male & 36-45 & Masters Degree & 3 \\
P8         &  Non-Binary & 18-25 & High School & 3 \\
P9         &  Female & 18-25 & Masters Degree & 2 \\
P10        &  Non-Binary & 26-35 & Masters Degree & 5 \\
P11        &  Male & 46+ & Undergraduate Degree & 5\\
P12        &  Male & 26-35 & Masters Degree & 4 \\
P13        &  Female & 46+ & Undergraduate Degree & 4\\
P14        &  Male & 18-25 & Undergraduate Degree & 2 \\
P15        &  Female & 18-25 & Masters Degree & 4\\
P16        &  Male & 46+ & Undergraduate Degree & 4\\
P17        &  Non-Binary & 26-35 & Undergraduate Degree & 2 \\
P18        &  Female & 46+ & Undergraduate Degree & 2\\
P19        &  Male & 46+ & Masters Degree & 5\\
P20        &  Male & 18-25 & High School & 3\\
P21        &  Male & 26-35 & Masters Degree & 2\\

\bottomrule
\end{tabular}
\label{tab:participants}
\end{table*}
}

\subsection{Participants and Recruitment}
We recruited participants through a screening survey distributed using our institution's mass email service.
We conducted interviews with 21 participants (see Table~\ref{tab:participants}) over a one-month time period from June 21st to July 25th, 2024. 
All interviews were performed in-person at a private location on our institution's campus, and consent was obtained from participants at the beginning of the interview.
Upon completion of the study, participants received \$30. 
The study was approved by our Institution's IRB. The length of each interview was between 48 to 82 minutes, averaging 64 minutes.

Over 849 people filled out the screening survey, with 831 being qualified for the interview.
Our screening survey asked respondents if they were comfortable with an in-person interview, their age, and to list what social media they currently use.
The exclusion criteria included respondents who were not comfortable with an in-person interview, below the age of 18, or did not use Instagram or TikTok.
Individuals who could not attend in person were excluded due to parts of the study requiring interaction with the machine learning models.
However, P18 was selected even though they did not actively use Instagram or TikTok.
We decided to include them in this study as they used Facebook and, therefore, indicated to us that they were familiar with the platform.
They also informed us, during the interview, that they wanted to get an Instagram to monitor their child's performance in sports.

In the survey, we collected information including how many hours a week participants used social media, whether they use social on their cellphones, their social media usage patterns (hours per week, platform of choice, commonly used features, etc.), and demographics. We then selected participants who used their phones for social media, balanced age, self-identified gender, tech-savvy-ness(~\cite{sudzina2015gender}), and their frequency of social media use.

To gauge tech savviness, we presented two statements: (1) ``People consider me to be tech savvy.'' and (2) ``I consider myself tech savvy.''
Individuals rated these questions on a 5-point Likert scale.

\label{sec:proto}
\subsection{Semi-Structured Interview Procedure}
\label{interview_protocal}
\begin{figure}[t]
    \centering
    \includegraphics[width=0.8\columnwidth]{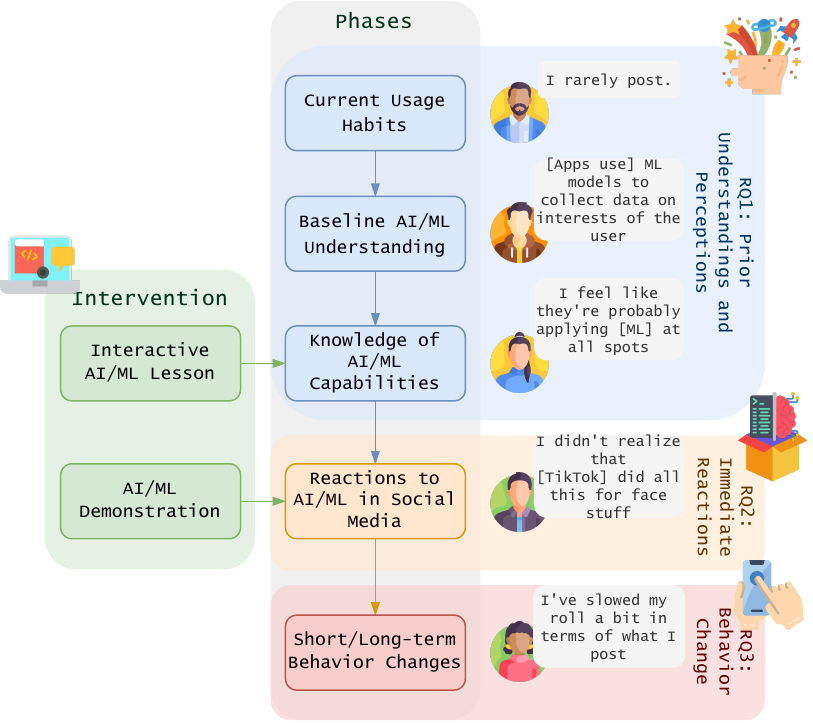}
    \caption{A high-level overview of our semi-structured interview, including example responses. The interview consists of five phases: (Phase 1) We explored participants' past interactions with their social media apps. (Phase 2) We identified participants' current understandings and assumptions towards AI/ML. (Phase 3) We conducted a learning activity, including a high-level example of computer vision, and discussed its capabilities. (Phase 4) We demonstrated the AI/ML models for TikTok and Instagram and observed their immediate reactions. (Phase 5) We surveyed short-term and long-term changes in participant behaviors. }
    \Description{The chart contains Phases, Interventions, and example quotations for each phase. The Phases are visually organized in a vertical column, with an additional vertical column for Interventions on the left. The example responses are introduced on the right of each phase. The key components are listed as follows. Phase 1: Current Usage Habits, and an example response is ``I rarely post''. Phase 2: Baseline AI/ML Understanding, and an example response is ``[Apps use] ML models to collect data on interactions of the user.'' Phase 3: Knowledge of AL/ML Capabilities, and an example response is ``I didn't realize that [TikTok] did all this for face stuff.'' Phase 4: Reactions to AI/ML in Social Media, and an example response is ``I've slowed my roll a bit in terms of what I post.'' The interventions, consisting of ``Interactive AI/ML Lesson'' and ``AI/ML Demonstration'', are linked to Phases 3 and 4. Three horizontal boxes representing the research questions overlap with the phases on the rightmost side. Phases 1-3 are in the range of RQ1: Prior Understandings and Perceptions; phase 4 is in the range of RQ2: Immediate Reactions; phase 5 is in the range of RQ3: Behavior Change. }
    \label{fig:methods_overview}
\end{figure}

As shown in Figure~\ref{fig:methods_overview}, our semi-structured interview included five phases that focus on the following topics.

\paragraph{Establishing Initial Understandings and Assumptions Towards AI/ML and Social Media}
We explored each participant's initial sentiments and perceptions regarding social media.
Specifically, we inquired about their favorite apps and why they use them.
We also asked about the content they consumed and their usage habits (posting, how they used social media, etc.).
Following the social media questions, we ask them about their understanding of ML.
These questions establish (1) if they had heard of ML before and (2) to explain what they think ML means.
If the participants had never heard of ML, we asked them if they had heard of AI.
They were then tasked with defining the difference if the participant had heard both ML and AI prior to the interview.

\paragraph{Defining Computer Vision and Clarifying Misconceptions About AI/ML} 
We transitioned to a learning activity to support users' understanding of general capabilities of AI/ML.
Using Google's Teachable Machine~\cite{carney2020teachable} we guided participants through a simple example of training and using a computer vision model.
To avoid common pitfalls from misconceptions created by machine teaching tools demonstrated by Hong et al.~\cite{hong2020crowdsourcing} we taught the participant ourselves avoiding an entirely automated process.
We designed a brief lesson where the participant trains a model that detects when they are in the frame and when they are not in frame.
To facilitate the process we provided props to the participant to test different classifications.
Once the new model was trained, we asked questions regarding their understanding of AI/ML based on the new content they practiced.
This phase allowed for each participant to engage in more in-depth conversations relating to AI/ML concepts.

\paragraph{Asking Participants ``When'' They Think AI/ML Happens in Social Media } 
We defined AI/ML expectations for social media applications in two ways: \textit{when} AI/ML should be used on their images and \textit{what} data should be observed from their images.
We defined four instances of when the model could be used: ``Opening Photo", ``Editing Photo", ``Face Filters", and ``After Posting".
Each scenario was demonstrated to the participant using Instagram on our study's Android phone.
For each instance, we asked participants when they believed AI/ML was happening and why.

\begin{figure}[t]
    \centering
    \includegraphics[width=0.8\columnwidth]{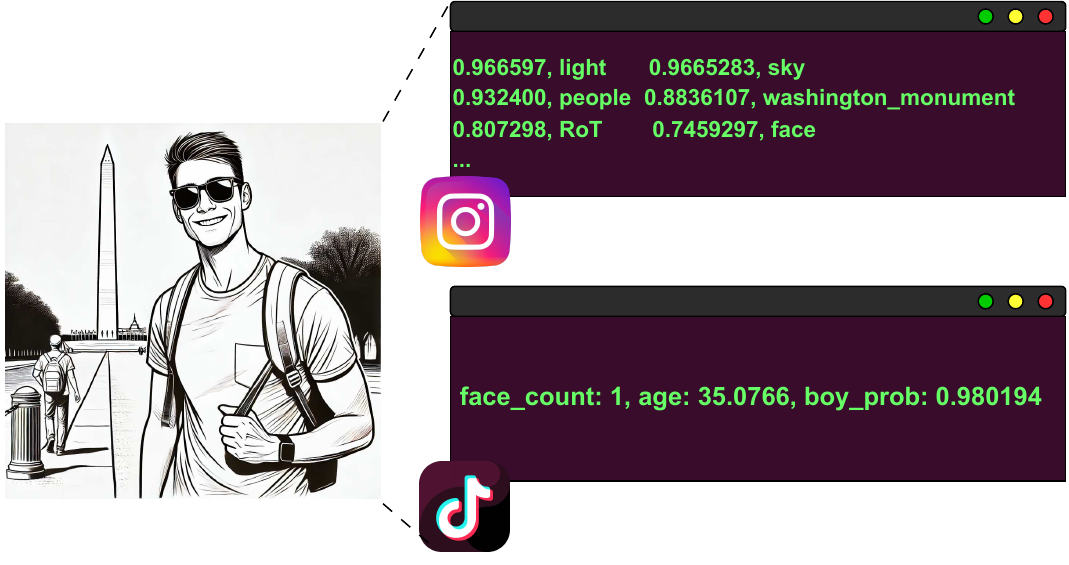}
    \caption{Example output that participants would observe during the interview.
            Instagram displays over 500 different decimals and concepts which are ordered from largest decimal to smallest.
            If the decimal value is closer to one than it Instagram perceives that concept in the image; if the value is closer to zero the concept is not in the image.
            TikTok observes age and gender age is an exact estimate and gender seems to be a probability.
            Gender is a probability called \textit{boy\_prob} when the decimal is closer to one it assumes a male gender identity if its closer to zero it assumes a female gender identity. Note that here we use a image generated by GPT-4o mini~\cite{gpt4omini} for illustrative purposes instead of the input picture from real people in respect of individual privacy.
            }
    \Description{The figure on the left shows a drawing of a man standing in front of a recognizable monument, the Washington Monument. On the right part, two separate data analyses from both Instagram (top) and TikTok (bottom) are linked to the left side. On Instagram Panel, it provides the probabilities for various objects detected in the image: 0.966597 for [light], 0.9665293 for [sky], 0.932400 for [people], 0.8836107 for [washington_monument], 0.807298 for [RoT], and 0.7459297 for [face]. On the TikTok Panel, the number of detected faces, the estimated age, and the probability of predicted gender. In this panel, the data shown are face_count: 1, age: 35.0766, and boy_prob: 0.980194.}
    \label{fig:outputs}
\end{figure}

\paragraph{AI/ML Model Demonstration and Participant Interaction} 
We demonstrated the authentic AI/ML models in real time. 
These models were adapted from the methods from West et al.~\cite{west2024picture} (for details see Sec~\ref{sec:model_extraction_details}).
We first showcased TikTok to the participant and followed by Instagram. 
The interviewer first showed how the AI/ML model works with their own face using the app.
We then asked the participant to observe the output (example output in Figure~\ref{fig:outputs}) and describe whether the model lived up to their expectations, if they were comfortable with the what they observed, and if they would've liked to know about this before using the application. 
The participant then explored the model output themselves.
While exploring the models, participants were informed that we were unaware of the purpose for the model outputs.

\paragraph{Assessing Short-Term and Long-Term Behavior Changes} 
We asked participants general questions comparing their experiences with the two applications.
During this process we asked about their feelings, reactions, and thoughts about what they have been shown.
We also asked them to decide which method of AI/ML they preferred and why.
Two-weeks after the last interview, we sent out a brief post-study reflections survey.
The survey inquired about if any behavior changes since the interview and/or if sentiment changes towards social media.
We sent the survey to all 21 participants and 17 responded.

\subsection{Extracting AI/ML Models from Instagram and TikTok }
\label{sec:model_extraction_details}
With the increase in technological abilities of mobile devices, there has been an industry-wide push to host local ML models.
Hosting ML models locally offers several benefits, such as increased user privacy, low latency, and algorithm personalization.
ML models of any modality are loaded using either a third-party library like Pytorch~\cite{pytorchPyTorch} or a custom ML library specific to the app.
Once loaded, ML models operate as they would on a server, existing as large blocks of memory performing inferences on input data points.
However, since the beginning of this migration, security researchers developed methods to extract, steal, and corrupt local ML models~\cite{sun2021mind, huang2021robustness, huang2022smart, cao2023cheating, deng2022understanding, west2024picture, ren2024demistify}.
To access ML models, three conditions must be met: (1) the researchers must have access to a rooted Android device, (2) the app containing the ML model must be statically analyzed, and (3) the app must be dynamically analyzed.
To fulfill the first condition, we acquired a rootable Pixel 4 device and followed a guide to root the device~\cite{xdaforumsGuideNovember}.
To accomplish the static and dynamic analysis requirements, we used the same methodology described in West et al.~\cite{west2024picture}.
We first decompiled the apps using JADX~\cite{githubGitHubSkylotjadx} and Ghidra~\cite{ghidrasreGhidra} to search for keywords that relate to machine learning.
Once relevant functions were identified, we then dynamically analyzed the apps from the functions of interest using Frida~\cite{fridaFridaWorldclass}.
However, for TikTok, the described method does not work due to its reliance on dynamically defined functions.
West et al. provide the function name~(\textit{MessageCenter.postMessage}) where the AI/ML output is displayed, which we use in this work.

\subsubsection{Our Understanding and Assumptions of The Models} 
Given that our method for accessing TikTok and Instagram's models was through reverse engineering, we have limited insight into how the models function, why, and what the information is collected for.
We found that TikTok's AI/ML model is built into the camera driver, meaning that all camera frames are directly handed to the model while the camera is open.
The model is \textbf{always} active, a filter does not need to be enabled, and we did not enable any filters when demonstrating the models to participants.
The data that we show participants are the result of the outputs being passed \textit{out} of the camera driver and into the Java layer through a \textit{callback} function.
The data is then written to an encrypted file.
Instagram's model executes when the user selects an image to upload as a Reel.
This selected image \textit{does not need to be uploaded}.
We found that the model only executes upon the selection of a photo.
We \textbf{cannot} assume the data's purpose due to technical limitations~(see Section~\ref{sec:tech_limits}).
We also checked Instagram and TikTok's privacy polices~\cite{tiktokPrivacyPolicy,instagramPrivacyPolicy} and could not confidently connect their privacy policy terms to either model we study in this work.
Thus, we decided \textit{not to inform} participants of our reverse engineering endeavors as it could lead to a bias. 
During our interviews, we informed participants that we were unaware of where the data went or what it was used for.

\subsection{Data Analysis}
To analyze the semi-structured interview using reflexive thematic analysis~\cite{braun2019reflecting, wilcox2023infrastructuring, mcdonald2019reliability} (RTA). 
RTA acknowledges the subjective and interpretive roles of researchers, viewing them as active contributors to meaning-making rather than objective coders. 
This approach was particularly beneficial given our team’s diverse expertise in cybersecurity, machine learning, and HCI; it allowed each researcher’s unique perspective to inform and shape the development of codes and themes. 
As a result, inter-rater reliability measures were not applicable within this reflexive framework~\cite{braun2021one}.
To transcribe the audio, we use the tool noScribe~\cite{noScribe}, which transcribes audio locally without third-party APIs.
All transcriptions are then verified by human reviewers.
Three researchers coded the first three interviews separately and had productive and iterative discussions pertaining to developed themes.
The researchers held periodic meetings to discuss codes, themes, and findings. The codes and themes were identified and developed in this process, and the disagreements were resolved collaboratively. 
The codebook was finalized by the sixth meeting, after which we continued discussing and analyzing our results.
All other transcripts were then coded using that codebook by a single researcher.
We also collected binary and open-ended answers about how participants' behavior changed in the post-interview survey. In the analysis, we clustered the participants into three categories based on their answers: changed behavior, did not change behavior, and planned to change behavior.  
The final themes and survey results are reported in Section~\ref{sec:Results}.

\section{Results}
\label{sec:Results}
We identified several recurring themes from our qualitative analysis, which we present in the following seven categories: (1) participant assumptions towards AI/ML and social media, (2) participants' usage behaviors, (3) perceptions around model deployment, (4) AI/ML privacy, transparency, and trust within social media, (5) comparisons between the two models, (6) data collection assumptions, and (7) reactions to the AI/ML models within social media.

\subsection{Participant Assumptions and Initial Perceptions on AI/ML and Social Media }
Among initial perceptions and understandings toward AI/ML and social media, we noticed three recurring themes highlighting participants' assumptions: (1) app interactions (likes, follows, searches, etc.) were assumed to be the main inputs to AI/ML for advertising and recommendation algorithms, (2) vision models were assumed to be used as a source to infer demographics in the context of social media, and (3) participants were unaware of the speed of modern computer vision models.

\subsubsection{The Purpose of AI/ML in Social Media is for Advertising and Recommendations}
\label{sec:gen_ai_purpose}
Our participants shared similar assumptions on how social media used information, mainly focusing on their content interests. 
They all believed that social media learned about them through their interactions within the app.
P4 explains 
\begin{quote}
\textit{``I think kind of the obvious [ML method] is they're using machine learning models to collect data on interests of the user. So understanding or collecting data on what they interact with, whether or not they like something, comment on something, share something, follow more pages that are related to a specific bucket of content.''}
\end{quote}
P20 expressed similar assumptions, stating that apps will probably look for words, topics, and accounts people interact with.
Participants assumed that data being observed by social media had the purpose of improving the experience (N=6), for profit (N=13), or just to collect data (N=3).
P3 explained their skepticism of free social media apps earn profits by collecting data that \textit{``is giving them profits in some way, shape or form.''}
Several participants assumed that \textit{user interactions} were collected to provide advertisements and/or recommendations all to increase profit (N=8).
We demonstrate that participants assumed \textit{user interactions} were the inputs to their advertising and recommendation algorithms.

\subsubsection{Assumed Purpose of Computer Vision Models in Social Media}
\label{sec:vision_model_purpose}
We asked participants, during the lesson, what they thought computer vision models could be used for within social media, and two main themes developed: (1) to offer features or improve algorithms (N=4) and (2) to infer more information about the user (N=15).
When asked how they thought a social media company could use a computer vision model, P9 assumed social media would want to find demographic data because they felt Instagram already \textit{``have [P9] pretty pinned down.''} 
Participants assumed that, for a successful personalization algorithm, social media apps would need to know more about the user.
P15, when asked about the purpose of computer vision models in social media, stated,
\begin{quote}
    \textit{Probably gender, age, location, geographic location. I think part of their algorithms or models would especially now be driven toward what would cause me to purchase things or what they should advertise to me. }
\end{quote}
P15's statement subtly implies that computer vision models may be able to identify several demographic indicators and that data would best be used for advertising.
Those who assumed that the models existed to improve their experience thought that the models did so by collecting demographic information or abstracting engagement visually.
P16 assumed that the data collection was necessary for the app as vision models could help everyone they explained, \textit{`` I mean, if [the models] helps man detect and helping people, I mean, that's beneficial.''}
P14 thought that Instagram's model could be used to detect explicit content, stating,
\begin{quote}
    \textit{``So yeah, like nudity is one of them up there. It's probably like, you know, you kind of have to like regulate that.''}
\end{quote}
P5 thought that vision models would be used to \textit{``understand how people are interacting and to keep them interacting if they have high interactions with one visual topic, for lack of a better word.''}
Both groups acknowledged and were able to differentiate between app interaction models and vision data.
Notably, because participants associated vision models with inferring demographics, most participants assumed that demographic estimation was why social media \textit{needed} vision models.

\subsubsection{The Model is Faster Than Participants Initially Assumed Possible}
\label{sec:model_capabilities_awareness}
Participants who assumed that their images were being analyzed by the applications were still surprised by how \textit{fast} the model collected data.
P10 described how fast TikTok was when they were just \textit{``on screen for like a second,''} and \textit{`` it was still able to pull up all this data.''}
P4 similarly expressed their surprise, stating 
\begin{quote}
\textit{``It is still a little bit intimidating and whatnot, but I wouldn't put it past these things to be able to do it this fast. I've just never really thought about it too much. But yeah, I mean, I guess it is still surprising. It's impressive.''}
\end{quote}
Overall, the model execution speed to some participants also felt invasive as they could not identify when it was happening.
This highlights that participants assumed that AI/ML could be detectable if it was being used.

\subsection{Participants' Usage Habits Influenced Their Reactions to the Models}
\label{sec:usage_habits}
We identified that participants' past usage habits on social media may create a bias that could shape their reactions to the AI/ML models.
Participants introduce three primary reasons for using social media: leisure, social connections, and professional purposes.
All participants (N=21) use social media for \textit{leisure}, with some (N=3) indicating that leisure is their only reason for use. 
For instance, P6 said \textit{``For general use, day to day, just for enjoyment. I doom scroll on TikTok quite a bit. ''}
Most of the participants (N=18) used social media for a combination of leisure, social connections, and/or professional purposes.
The second most common reason was to maintain \textit{social connections} (N=16): P12 uses social media to keep up with friends and family, explaining, 
\begin{quote}
\textit{`` I come from [foreign country]...I wouldn't say it's a huge following. It's just that I've made friends there in this community there.''}
\end{quote}
Lastly, a few (N=3) participants use social media for a \textit{professional purpose}, such as P5, who has a social media to advertise their business, \textit{``primarily about [their] art practice,''} despite their rare posting.
Those who use social media as a source of leisure interacted with the camera \textit{less} than those who utilized the apps for social connections and professionalism.
The difference in participant \textit{interactions with the camera will influence their short and long-term reactions} to the models.

\subsection{Participant Perceptions Towards AI/ML Deployment}
\label{sec:deployment}
We explored participants' perceptions of when AI/ML should be utilized by social media applications.
We noticed that participants disagreed on \textit{when} AI/ML may take place.
\subsubsection{When AI/ML Takes Place Is Not Clear to Participants}
\label{sec:when_ai}
In the third phase of our interview, detailed in Section~\ref{interview_protocal}, we presented four options where AI/ML might take place for participants to choose from: (1) ``Opening Photo'', (2) ``Editing Photo'', (3) ``Face Filters'', and (4)``After Posting''.
We noticed \textit{disagreement} amongst participants for ``when'' they assumed AI/ML occurred on social media apps.
Half of our participants considered \textit{AI/ML happens in all presented scenarios} (N=10), and the remaining participants (N=11) chose \textit{a subset of scenarios they thought were likely}.
Among those participants in the second group, only one participant (P16) selected ``Opening Photo''.
Some participants in this group were weary of data being collected during the ``opening photo'' stage, P7 proclaimed \textit{``Because opening the photo is in my device, it's still not posted and they won't receive the data from me.''}
Participants who selected all four options felt that the ``Opening Photo'' stage was unlikely. P3 states that they \textit{``don't necessarily know about `opening the camera.'''} and feel like the apps are \textit{``probably applying at all spots.''}
P5 argues that this confusion correlated with a lack of transparency: 
\begin{quote}
\textit{``I know [AI/ML is] there. Google sends me like `oh six years ago', right? Like `oh and this themed slide shows of'...which are like cat pictures where they're all together...they're clearly scanning and watching what is in my photo album and so, it's like, you know it's there.''}
\end{quote}
Overall, we identified that \textit{participants were confused on when AI/ML actually happened} which, some participants argued, that the lack of transparency caused the confusion.

\subsection{Privacy, Transparency, and Trust in Social Media Applications}
\label{sec:priv_transparency}
We found that lacking trust played a role in several assumptions toward data usage and local data processing.
This lack of trust stems from social media's inferred business model: user data.
Participants assumed that the outputs from the models are likely used for some underlying algorithm to further profits.
\subsubsection{Participant Trust Influenced Assumptions Towards Local Processing}
\label{sec:trust}
When we informed participants that we did not know what happens to the model outputs, we observed that several individuals (N=11) expressed two different reactions: (1) locally processing the data \textit{would not} change their attitudes towards the models (N=6) and (2) local data processing \textit{would change} their attitudes (N=5).
P20 expressed that they want to know because \textit{``it's still using the technology that I own, because it's basing it off of the phone. It's using it in a way that I did not consent to, nor unless I'm lucky, I don't know what's happening.''}
But when local processing was further explained to P20, they stated, \textit{``[local processing] makes it a little less malicious than I was perhaps making it out to be.''}
P20's malice was a result of their perceptions of the data's purpose, stating they assumed \textit{``[social media apps would] likely just logging demographics and then maybe, you know, looking into like what they could boost based off of those demographics to.''}
Even when understanding that the readings would never leave the phone, they assumed that it would still be used for advertising locally.
These feelings stem from a lack of \textit{trust} in social media apps. When asked if P4's opinion of Instagram would change if the data did not leave the phone, they stated,
\begin{quote}
    \textit{``So unless I know, like, I have definitive proof of knowing where that goes, and I just assume the worst. I'm a skeptic in that way. And so, you know, Instagram and TikTok, they could tell me that that's on my phone and on my local server or whatever, but I tend to not believe that.''}
\end{quote}
P4 expresses a lack of trust towards the applications due to the monetary value of their data, stating, \textit{``So part of me says it would be good for them just to delete that data and not have it be out there. But I know that that's not their business model.''}
P3 thought the data \textit{must} be collected because \textit{``Instagram is a free, free thing. So where are they getting the money from? So obviously, right, they're collecting data, and that data is giving them profits in some way, shape or form.''}
P3 and P4 both assumed that their data had monetary value and thus assumed that it would be used or collected.
We identified that participants' lack of trust in social media influenced their opinions towards the apps processing local data.
Interestingly, participants' lack of trust is \textit{not} directed at the AI/ML models themselves but rather the applications deploying them.

On the other hand, P7 stated that if they knew their information did not leave the device, they would be fine with AI/ML analyzing their data.
P7 \textit{``assumed that social media is a business''} and our data was how they made profits, but \textbf{they must follow the law}.
When asked if P7 would implement their hypothetical algorithm designed for social media, they stated,
\begin{quote}
    \textit{``Ethically I wouldn't but... And legally I would... If it's not legally allowed to do, I wouldn't.''}
\end{quote}
P14 expressed a similar sentiment by assuming that the models are likely disclosed in the \textit{``terms of service''} without reading it.
While P7 and P14 put their trust in the law, other participants felt agnostic about local data processing because they didn't know where it went.
P12 said that their data had \textit{``so much potential''} and thus did not full dismiss the possibility that their data was being collected.
These results imply that only locally processed data may not raise trust in social media applications due to the assumed business model.
Those who did trust local processing only did so when the trust was built on the law.

\subsubsection{Participants Disagreed About How Much Transparency Should be Given to Users}
\label{sec:transparency_comfort}
All participants wanted more transparency about AI/ML models used on their devices.
However, there was disagreement about how much data apps should be transparent about.
Participants exhibited one of three themes: (1) apps should detail as much information as possible (N=12), (2) apps should only provide limited information (N=5), and (3) the participants felt they already knew about data collection; thus transparency felt unnecessary (N=3).
Most participants expressed curiosity about the models' purpose after being exposed to them.
One example, P13, stated that \textit{``Not knowing what the purpose is for makes it harder to be accepting that it's happening.''}
Others expressed a want to \textit{control} the data.
P5 explained that they, \textit{``would like control and not be able to give it information that I don't want to give it,''} when thinking about how their data could be used for advertising.
However, P5 later acknowledges that it may render their algorithm useless, stating, 
\begin{quote}
\textit{``But then what good is the algorithm? And what are we using this algorithm for?''}    
\end{quote}
P5 expresses mixed feelings as they recognize that control could halt social media from learning about them but, at the same time, would hinder their algorithmic experience.
Some participants wanted to know, but they did not want all of the information we provided.
P11, for example, thought there was too much detail, especially regarding Instagram's model.
P9 and P13 worried about the potential societal impact when revealing the outputs to the models.
P13 assumed people would get side-tracked by the model, which would create a distraction.
P9 assumed that people would attempt to optimize their scores, and it would drastically harm the next generation, stating,
\begin{quote}
    \textit{``I mean, like people would be like, why is it my face 0.9 compared to Becky's face to one? Now I don't have as many followers because the score isn't higher. And how can I get it higher? And yeah. I think it could be detrimental for sure.''}
\end{quote}
P9 reflected that users may \textit{define} their self-worth by the outputs from the model.
These assumed risks of increasing transparency are a direct result of the context of social media.
The final notable opinion is that the participants knew this was happening and, therefore, felt it did not apply to them.
For example, P16 said that they \textit{``kinda of figured''} something like this was already happening and didn't see a need to be told because, to them, the fact the model existed was obvious.
Overall, we identified that participants expressed \textit{multiple levels of comfort towards transparency.}

\subsubsection{Some Participants Did Not Expect the Camera to Analyze Them}
\label{sec:camera_look_at_them}
Participants were confused about ``when'' a model was executed.
P8 expressed surprise when the model was active, \textit{``I didn't realize that [TikTok] did all this kind of stuff for face stuff. I don't know. I just thought the camera just looked at you.''}
P1 similarly reflected on how uncomfortable it felt knowing they had opened TikTok's camera in the shower before,
\begin{quote}
\textit{``That's quite surprising. Sometimes when I take a shower, I sometimes play [with my] cell phone. I mean I use my cell phone when I take a shower. Well, now that's... I'm naked''} 
\end{quote}
These examples demonstrate the vulnerability and risk associated with AI/ML models use of passive camera stream.

\subsection{Participants Compared Instagram and TikTok's Model Implementations }
\label{sec:comparisons}
AI/ML models deployed by Instagram and TikTok analyze user images (Instagram) and camera frames (TikTok) on user mobile phones.
Participants exhibited various reactions after experiencing how the TikTok and Instagram models operate: their reactions centered around the models' \textit{inputs}, which is how images are provided to the model, and \textit{outputs}, referring to the models' analysis of the given images.
Participants shared that TikTok's model felt non-consensual because of how the images were delivered to the model.
However, participants were uncomfortable with the amount of data that Instagram observed. 
We found that ten people preferred TikTok's model, while only seven people preferred Instagram's.

\subsubsection{Comparisons Toward Model Input Methodology}
\label{sec:input_comp}

\paragraph{Participants Felt That TikTok's Model Was Creepy}
Due to the fact that TikTok's model was active while the camera was open and undetectable to participants, they felt it lacked consent to analyze their images.
When asked how they felt about local image analysis on their data, P20 angrily emphasized their disappointment of having \textit{``no way of knowing what it's picking up on, what it's storing...and what it's doing without [their] consent?!''}
P17 reflects on TikTok's model output, considering that \textit{``plenty of people might accidentally open their TikTok camera''} without \textit{``thinking about giving data in that moment.''}
Participants felt that because TikTok did not notify users about their data being collected, they felt TikTok was \textit{creepy}, especially to P13, who stated, \textit{``So TikTok, your camera is open. You haven't even like recorded or posted and it's doing that? Oh, yeah. That's that's creepy!''}
P19 felt that TikTok was \textit{````Big Brother-ish''}
\begin{quote}
\textit{`` feels more like a place that nobody else deserves to be. Just me. Almost like somebody hacking into and watching you through your camera at home and you don't know it.''}
\end{quote}
Overall, participants expressed negative perceptions upon learning how TikTok's model functions due to the lack of consent for data collection, and the limited transparency regarding their practices further exacerbated the participants' feeling `creepy' and invasive.

\paragraph{Instagram's Model Felt Less Creepy}
All participants had a notably better perception of Instagram's model implementation, and they considered it \textit{less} creepy than TikTok's. 
P20 thought \textit{``Instagram model is more consensual because you have to take an image or look at an image that's already on your phone to activate the model.''}
This suggests that with Instagram's model, there was a sense of \textit{control}, which was missing from TikTok's model.
P18 explained, 
\begin{quote}
\textit{``Instagram is being a little more discerning and letting you do some selection before it runs through that filter and waits to see what you take a picture of before it turns this on.''}
\end{quote}
These feelings were strong enough for several participants (N=7) to prefer Instagram over TikTok because of the extra perceived control. 
P18 thinks \textit{``TikTok seems more aggressive, but it doesn't seem like as radical a difference [between Instagram and TikTok].''}
When asked to clarify what they meant by `aggressive,' P18 refers to TikTok's attempt to assess \textit{``who you are and what you might buy the second you turn on the camera.''}
This extra control on Instagram soothed some participants as they now knew \textit{when} the AI/ML could happen, and they felt in control. 

\subsubsection{Model Outputs are Confusing and Misaligned with Expectations}
\label{sec:output_comp}
The negative reactions to Instagram were shaped around the magnitude of data presented as part of the model's output.
Participants felt the data was \textit{random} as the sheer amount of data was difficult to comprehend.
Some participants (N=8) tried to assign \textit{assumed meaning} to the data by connecting their folk theories to the new data.
Participants were \textit{confused} as to why Instagram selected the values to assign to their images.

\paragraph{Instagram's Data was Too Confusing}
The majority of participants were confused by the outputs.
P19 described the labels as a \textit{``drunken mistake''} and found \textit{``it very fascinating in several different ways.''}
The sheer amount of data lead most, who did not assign meaning to the labels (N=8), to believe it was useless. 
P15 explains why they do not feel uncomfortable necessarily as \textit{``this doesn't mean anything.''}
Whereas some participants (N=2) associated the data with negativity P13 summarizes how others might feel, by stating 
\begin{quote}
    
\textit{``I'm still kind of neutral to negative about it because there's so many words that they are looking that don't apply at all. And so then again, I'm like, what's happening with this information? And then like even saying coming up with percentages that comes out of this many people do this. It's like, well, I'm not at a beach. And I don't know things like that.''}
\end{quote}
P15 argues that the confusion is a product of not grasping where the data fits in an algorithm, and they \textit{``don't really have an explanation for why''}, when comparing the algorithms of TikTok and Instagram.
TikTok's model appears to P15 that it infers attributes like gender and age which \textit{``are two driving factors of how people interact with media in general, not just social media;''} while Instagram's model feels like \textit{``they're trying to understand more about what's in the picture.''}
These reactions are likely rooted in the participants' initial understanding of the data's likely purpose.

\paragraph{Instagram's Data Felt Random}
The most common reaction was that Instagram's data felt `random' (N=19), as described by P3 who thinks \textit{``it is surprising in the randomness of these categories.''}
Participants also noted that the data was confusing, as put by P12, 
\begin{quote}
\textit{``I guess my question is why... I mean, there's more than 500 things in the world, right? So why these 500 categories?''}
\end{quote}
To the participants, this information was difficult to understand as P15 feels \textit{``like they're trying to understand more about what's in the picture, but [they] don't really have an explanation for `why'.''}
This reaction highlights a potential flaw with being transparent.
When revealing the model outputs, the participants felt the data was \textit{random}.
The data feeling random lead to several assumptions about the data's purpose or assuming that the data had no purpose.

\subsection{Participants Were More Likely to Accept Data Collection Practices When They Understood the Data}
\label{sec:data_collection}
Participants were often able to make sense of TikTok's data.
Some did not like the way the model was used without explicit consent but felt that they \textit{understood} why the data was being analyzed (N=8).
P20 immediately assumed that TikTok's data collection, including demographics, was for advertisers or to \textit{``farm engagement''}
More participants preferred TikTok (N=10) over Instagram. 
This is likely due to TikTok collecting less data and participants feeling like they understood what the data was used for.
When asked which model they were less comfortable with, P8 refers to Instagram as \textit{``it seems like there's all these different things that they're trying to categorize, which is a bit weird.''}
P4 similarly contributes to this line of thought after comparing both applications
\begin{quote}
\textit{``I would be more likely to use TikTok because I think that it's collecting far less data and maybe, maybe less harmless data. That might be not a true statement. But yeah, just based on what you've seen. Yeah, I'll say because of less, less data.''}
\end{quote}
Overall, the lack of transparency towards the model likely left people confused and speculating about the purpose of the data.
Which, in turn, leads to further algorithmic theories or confusion about what this data may be used for.

\subsection{AI/ML Model Reactions and How They Influenced Behaviors}
\label{sec:behaviors}
We interviewed participants to see whether they thought their behavior would change in response to the intervention during and 2 weeks after the interview. 
17 out of 21 participants replied to the post-study survey we sent to inquire about long-term behavior changes.
We refer to the participants' answers immediately post-intervention as \textit{short-term} changes, and we refer to the answers to the post-interview survey as \textit{long-term} changes.
We found that 6 participants expressed short-term change, and 8 expressed long-term change.
Notably, 4 of these 8 participants had stated that they did not plan to make any long-term change.

\subsubsection{Participants Inferred Meaning Toward Instagram's Data}
\label{sec:meaning_of_data}
Participants expressed one of two reactions with regard to Instagram's output: (1) they could infer use cases for the data (N=8), or (2) the data did not make sense (N=11).
Those who gave meaning to the labels had mixed reactions based on their assumed purpose or consequences.
For example, P10 was uncomfortable with the concepts related to location, \textit{``even if [they] don't post my face specifically.''}
Similarly, P7 does not \textit{``want any company to get location from [them] just with a picture.''}
These negative reactions focused on labels that they \textit{personally} did not like.
However, there were also positive feelings toward the potential uses of the data. 
For example, 
\begin{quote}
\textit{``Like even though some of [the concepts] are goofy to me. They seem less. They seem more objective in a way that is less dangerous to get wrong.''}
\end{quote}
P18 assumed that because the concepts were general if the AI/ML model were to get them wrong, there would be fewer consequences compared to TikTok.
Due to the lack of transparency, participants were confused as to what the purpose of the data was.

\subsubsection{Models Validated Prior Assumptions}
\label{sec:vaild_assumptions}
After seeing the models, P8 felt vindicated in their decision to delete TikTok because they felt like \textit{``it was way too personal.''}
Participants often referred to the models as ``eye-opening,'' stating that some prior assumptions before the interview were validated when observing the models.
P13 felt like their thoughts have been confirmed after exposure stating \textit{``this is like in your face. This is what's happening.''}
P9 shared a similar sentiment and already assumed that \textit{``everything [they] hear about machine learning and data collection''} was happening. 
After an \textit{``eye-opening''} experience of seeing the model, P11 discussed further assumptions if it operates at \textit{``camera level''} or \textit{``transmit level''} and combines data \textit{``in a million different ways.''}
These responses indicate that our intervention offered some context to participants' prior perceptions about AI/ML use on social media.%
P5 speculated how Instagram's model is used to keep people on the app, 
\begin{quote}
\textit{`` [The model is] useful for [Instagram] as far as their goals of keeping people on the app and how they see their end for capturing the data that would be useful for them as far as how people interact or how people use this information and being able to mine data of who's interested in what and then being able to sell that data or, I don't know, I don't know. Big Brother's watching''}
\end{quote}
Some participants referred to data as `profits'. P7 thinks social media companies \textit{``have to make profit for the business to keep running and dominate the one over the other.''}
P11 similarly reflected \textit{``if they're looking for these things, it's because there's profit possibility.''}
P12 stated \textit{``There is a reason all these apps are free.''} 
Further, P12 argued that it is unethical to collect this information, or \textit{``it should be clearly listed how the information should be used''} at least.
These quotes emphasize that participants' perceptions of the social media apps relates to the business model of the apps. %

\subsubsection{Participants’ Short-Term Reactions to AI/ML Models were Shaped Based on Their Personal Values and Cost-Benefit Trade Offs}
\label{sec:short_term_behaviors}
Short-term reactions to the models tended to centralize around if the model \textit{impacted them} (N=6).
When asked if location data was a concern to them, P7 replied \textit{``Definitely...because [they] don't want others to know about [their] location and where [they] move, where [they] stay, where [they] travel. ''}
After being asked if they see themselves changing behaviors post model exposure, P7 stated 
\begin{quote}
\textit{``The social media, like Instagram, posting pictures, starting a record, or just opening a camera. I'll be more conscious before even opening the app.''}
\end{quote}
Due to P7 personal privacy considerations, they \textit{did not} like that Instagram was estimating location based on the image before posting.
P10, also uncomfortable with the location being estimated, shared P7's sentiment.
P7 and P10 both stated that in the long term, their behavior did change P10 states that they \textit{``will be posting less identifying information on social media,''} although they \textit{``will still scroll through the timeline.''}
However, other participants felt like the models didn't apply to them (N=15).
P3 assumed that their privacy is not important, stating 
\begin{quote}
\textit{``At the same point, it's like I'm not at a military base. You know, I'm not in a foreign country. I'm not having state secrets behind me. So it doesn't really matter that much to me that they're collecting all of this.''}
\end{quote}
Whereas, when asked if they would change their behavior, P5 says no \textit{``because [they] don't open up [their] camera''} and they have not been posting.
P9 offers similar reasoning for not changing their behavior as they have already been carefully and only occasionally posting.
What P5, P3, and P9 demonstrate is that their current habits are \textit{unaffected}.
Participants who felt unaffected by the models felt less of a reason to stop using the apps.
This was most common amongst those who use the app for leisure (N=3).
P9 stated they used social media 
\begin{quote}
    
\textit{``for the entertainment aspect for sure. And this is like the distraction in my life. And I guess I'll just keep [using Instagram].''}
\end{quote}
Overall, habitual usage may effect short-term reactions to the apps.

\subsubsection{Some Participants Experienced a Slight Decline in Social Media Usage After the Study}
\label{sec:long_term_behaviors}
While six participants reported short-term change, eight reported long-term change.
Notably, out of those six, four reported long-term change, and one reported that they did not change after the interview (P20).
This implies that four of the long term behavior changes stemmed from those who originally said they felt the models did not apply to them.
P5, one of the participants who originally said they would not change, found that 
\begin{quote}
\textit{``in addition to using Instagram less frequently I have been using Reddit more instead. It still has not changed the fact that I do not use TikTok, though I am even less keen on opening TikTok links that friends share with me.''}
\end{quote}
P19 slowed their usage of Instagram particularly, their \textit{``usual place for more personal photos.''}
Whereas, P4 mentioned their actions to \textit{``to reduce permissions for these apps to access personal data''} along with their lower usage frequency of social media.
P4, P5, and P19 felt, during the interview, that their behaviors likely would not change, but all of them felt \textit{negatively} about the models.
P19 described their feelings about their usage during the interview as 
\begin{quote}
\textit{``Sometimes I'm able to stop myself and say, is there anything positive about this? And if there's not, don't do it. I can turn that into Instagram, saying, is this information that doesn't need to be out there? And I am protective and private in that way. Unlike some of the other people I know who just put... I think it'll make me more cautious.''}
\end{quote}
P4, on the other hand, understood why their friend's parents wanted their faces blurred in social media posts, implying that they now understood why the parents wanted to be blurred. 
However, both P4 and P19 expressed hesitance toward changing behaviors, but both reported that they now understood the caution that others took.
During the interview, when asked if they felt their behaviors would change, P4 stated,
\begin{quote}
    \textit{``But I'm kind of in this spot right now where I feel like I've already been compromised, per se. And I don't know if it would make too much of a difference if I stop now.''}
\end{quote}
This indicates that negative feelings from the intervention may have influenced participants to change behaviors in the long term.

\subsubsection{Model Outputs Felt Personally Offensive}
\label{sec:offensive_outs}
Participants' personal interpretations to the outputs from the models influenced their reactions and behaviors toward AI/ML models.
P18 angrily expressed distain towards TikTok's gender estimation, 
\begin{quote}
\textit{``Because I'm the parent of a generation of kids here in [location], where the assumption is that you just don't make, they don't make assumptions about gender, period. ...  And I am the [parent] of a non-binary child, and so the [protector] kicks in and really pisses me off when people [misgender them].''} 
\end{quote}
Similarly, P13 did not like TikTok estimating the age of children, claiming that it reminded them of ``human trafficking''. 
However, out of the two participants \textit{only P13} expressed a change in their long-term behavior.
P13 stated in their post-study survey: 
\begin{quote}
\textit{``I immediately went into TikTok and changed my settings to deny access to my camera and mic. I use Instagram much less and am more conscious about what I consider posting.''}
\end{quote}
They passed this information onto their young adult children:
\begin{quote}
    \textit{``I'm definitely more cautious and wary of social media apps in general. I’ve shared what I learned with my young adult [children] too.''}
\end{quote}
Notably, P13 was an avid user of both TikTok and Instagram.
P18, on the other hand, did not use TikTok or Instagram and reaffirms the cautiousness toward using Instagram \textit{``which [they] do expect to use in the future.''}
Arguably, this difference between participants may be due to their personal experiences of feeling impacted by the model.

\section{Discussion}
In this work, we explored social media users' perceptions and reactions toward transparency in AI/ML models deployed by social media. 
Our work is motivated by the limited transparency that social media applications offer to their users regarding their data collection and use practices for AI/ML models. 
Our findings offer insights into ethical and practical opportunities to improve the transparency of AI/ML models. 
They also highlight several key issues that need further discussion, including integrating transparency into social media applications for their AI/ML models.

\subsection{Summary of Findings}
Several of the identified themes address our research questions.
We found that participants shared the perception that social media use AI/ML for algorithm personalization and/or demographic identification (RQ1).
We also demonstrated reactions from participants, comparing both models and their unique positive and negative comments tied to each (RQ2).
Finally, we identified that participants' self-reported behavior changes were influenced by how they used the app and/or personal reactions to the model (RQ3).

\paragraph{RQ1:}
Participants assumed that AI/ML is used for personalization and recommendation algorithms within social media~(Section~\ref{sec:gen_ai_purpose}).
In their general discussions around AI/ML, participants rarely discussed vision models.
However, when directly exposed to vision models, participants tended to assume that social media would implement these models to \textit{collect demographic information}~(Section~\ref{sec:vision_model_purpose}). 
While all participants expressed similar assumptions towards the purpose of the AI/ML models, their opinions varied on whether they were bothered by data being processed locally~(Section~\ref{sec:trust}).
We noticed a prevalent assumption from participants that social media companies \textit{need} to be profitable and users' recommendation and personalization algorithms are the source of profit~(Section~\ref{sec:priv_transparency}).

\paragraph{RQ2:}
Based on the prior assumptions of AI/ML in social media, participants assumed that the models were attempting to collect data even when explicitly told that the data may not be collected at all.
After revealing both models to participants, we noticed that participants expressed unique positives and negatives towards each. %
TikTok was criticized for their model always being active with no indication to the user~(Section~\ref{sec:input_comp} \& Section~\ref{sec:deployment}).
However, participants noted that the amount of data about the user from TikTok's model was noticeably smaller compared to Instagram~(Section~\ref{sec:output_comp}).
Participants were positive about how Instagram's model was executed (i.e., active only upon selection of a photo) but was critical towards the amount of data within the model's output~(Section~\ref{sec:output_comp}).
Our analysis revealed that TikTok's model aligned more with prior participant assumptions~(Section ~\ref{sec:data_collection}), but Instagram's outputs confused participants~(Section~\ref{sec:output_comp}).
The model outputs from Instagram contained noticeably fewer demographic concepts, which confused participants as the data's purpose was not clear.

\paragraph{RQ3:}
After seeing the models, some participants self-reported that they noticed small behavior changes, whereas others continued normal usage~(Section~\ref{sec:behaviors}).
Most participants stated that knowing about the models \textit{would not} change their behaviors as most participants did not actively use the camera for it to concern them~(Section~\ref{sec:short_term_behaviors}).
We noticed that participant usage habits had a major influence on participant reflection~(Section~\ref{sec:usage_habits}).
However, participants who valued their privacy to a higher degree did express that they aimed to change their habits after the interview.
Our post-interview survey illustrated that more participants \textit{reported long term change} than in the immediate reactions~(Section~\ref{sec:long_term_behaviors}).

\subsection{Integrating AI/ML Transparency in Social Media Applications }
A general theme found common across all participants is the lack of transparency of how and when social media apps employ local AI/ML. In the following, we discuss how user expectations did not match app behaviors. Then, we identified three transparency preferences discussed by participants. 
Finally, we explore a future research agenda based on our findings.

\subsubsection{Expectations}
Our study showed that participants did not know that their data was being analyzed and were unaware of the consequences of said analysis. 
Since users were unaware of the possibility that AI/ML could happen to their local images, they did not realize that simple actions such as selecting an image (and deciding not to post it) or opening the camera through the app (accidentally or on purpose) resulted in their data being analyzed. Participants in our study assumed their personal data was analyzed \textit{only after} an image was uploaded. This misconception is not the fault of the user, given that prior to recent developments~\cite{wang2018deep,dai2020state}, using a model on a server was the primary method for image analysis.   
Moreover, we observed that individuals \textit{do not necessarily understand the consequences} of the models.
Participants often expressed confusion about Instagram's model and had trouble connecting the model's output to any familiar functionality or use case.
This limited understanding may cause users to be unable to comprehend the consequences of providing their data.
How large businesses should collect data is still a debated topic among researchers~\cite{bhimani2015exploring,wiener2020big,koukouvinou2024ai}.

\subsubsection{Preferences}
Participants clearly articulated their expectations of needing more transparency for how and why AI/ML in social media apps process their data. In particular, we identified three preferences. First, some participants wanted to know more about the employed models in terms of inputs (Section~\ref{sec:input_comp}), outputs (Section~\ref{sec:output_comp}), and purposes (Section~\ref{sec:meaning_of_data}). Second, some participants wanted to know \textit{when} the models are being applied (Section~\ref{sec:camera_look_at_them}). Third, participants expressed interest in controlling or changing how models process their data (Section~\ref{sec:transparency_comfort}).

\subsubsection{Research Directions}
We propose three possible avenues for future research to improve transparency.

\paragraph{Install-time Transparency}

Our study emphasizes the important need to be transparent about \textit{when} data is analyzed, \textit{what} a model's inputs and outputs are, and the \textit{purpose} of a model.
At \textit{minimum}, participants thought that social media apps should provide notice about model inputs, outputs, and purposes (Section~\ref{sec:transparency_comfort}).
A potential research direction could be to integrate AI/ML information into existing install-time transparency solutions (privacy policies, privacy dashboards~\cite{priviawareLee}, privacy nutrition labels, etc.). 
However, we foresee three main challenges with integrating AI/ML transparency into existing frameworks: (1) what modality is best for social media users, (2) portraying AI/ML transparency in an accessible way, and (3) providers deploying AI/ML transparency mechanisms.

Some participants indicated they assumed the information about AI/ML they saw was in the apps' privacy policies~\cite{instagramMetaPrivacy,tiktokPrivacyPolicy}. However, users likely do not read privacy policies~\cite{steinfeld2016agree} and those who do must parse inaccessible legal language which may confuse users further~\cite{ibdah2021should,vu2007users}. Further, some participants claimed they would not trust any transparency mechanism presented by Instagram or TikTok. This raises an interesting research direction for social media AI/ML transparency: what modality would satisfy users of social media? Due to the trust issues unique to social media, existing solutions may require \textit{more transparency} than currently offered by existing frameworks~\cite{windlInvestigating,priviawareLee}.

To address the second challenge, we propose that future work explore user-friendly FactSheets~\cite{arnold2019factsheets} from the XAI community.
FactSheets are designed like a schematic for AI/ML containing details about the deployed model, such as the inputs, frequency of inference, outputs, performance test results, and purposes.
They are designed to be iteratively updated with the model and contain details about AI/ML risks such as algorithmic consequences of a wrongful prediction. 
However, FactSheets provide too much information as they are designed for developers of AI/ML.
The amount of information within FactSheets could lead to security risks~\cite{hu2021artificial} and would likely confuse users further.
Striking a balance between relevant information and accessibility is crucial for this hypothetical framework.
To address accessibility we argue that providing users with tools that allow for them to interact with the AI/ML could be beneficial~\cite{kim2023help} and/or explaining AI/ML decisions using human rationale~\cite{kim2023help}.

Our third challenge is how to encourage social media apps to implement the AI/ML transparency mechanisms.
For example, prior work has shown that users have a positive attitude towards Data Safety Labels~\cite{kelley2009nutrition} \textit{but} developers have voiced concerns and challenges during implementation~\cite{li2022understanding}.
AI/ML processes can be \textit{complex and non-deterministic}.
Developers may not know how a complex model would react with a given input.
For example, OpenAI did not foresee lawyers using their models to write legal documents and for their model to hallucinate fake cases~\cite{forbesLawyerUsed}.
Thus, determining the risks of a complex AI/ML model may not be possible.
We need to explore how to properly convey this lack of determinism to users especially as it pertains to their private data.

\paragraph{Run-time Transparency}
Participants were interested in an additional level of transparency: being notified at the time when AI/ML is active. In particular, they expressed interest in knowing \textit{when the models were active} to avoid them. 
Mobile phones offer operating system-level notifications upon access to a camera or microphone.
We suggest future works explore designing real-time AI/ML indicators to signify to the user that AI/ML is currently processing their data on the device.

An associated challenge is how much information this notification mechanism can provide the user. One option is to limit the notification mechanism to inform the user of a specific app engaging in AI/ML. Another option is to highlight the type of input data to the model, perhaps using different icons to represent data types. The notification can display additional information in an expandable interface, including the outputs of the model and potentially the purposes.

These notification mechanisms would require collaboration between a mobile platform and the app developers. As opposed to hardware resources, such as the microphone and the camera, AI/ML is an internal app resource, and it is unclear how the operating system can mediate access to the model. As a result, the app has to notify the operating system, and subsequently the user, that AI/ML processing is taking place.

\paragraph{User Control of Model Outputs }

As mentioned earlier (Section~\ref{sec:transparency_comfort}), some participants felt uncomfortable with the models' observations of their data. Several participants expressed interest in having the option to correct or change the model output. Our participants' preferences present a new challenge to the HCI community: designing user-level controls for local AI/ML models within social media. Such controls would ensure users could still interact with the app without consuming undesirable model outputs. Toward that end, we foresee an interface that enables users to (1) \textit{select} which concepts they want the app to process and (2) change the scores of concepts they find inaccurate.

Providing that level of detail and control comes with many challenges. The first challenge relates to implementation, which, similar to the notification dot, requires collaboration between the mobile platform and the app developer. There are also potential intellectual property issues when revealing model outputs. The second challenge is that models can be complex (e.g., Instagram), providing overwhelming information to a user. Addressing this challenge requires careful design of usable and informative interfaces towards what a user change could mean for their experience. The third challenge, brought by some study participants, is that revealing model outputs would affect users negatively because they would feel the need to optimize their scores. We foresee a complex relationship between trust and increased transparency. On the one hand, revealing \textit{too much} detail to the user can weaken trust~\cite{ehsan2021expanding}. On the other hand, transparency mechanisms that explain the purposes of potentially intrusive data processing in the context of safety and utility might increase trust.

Finally, some scores \textit{should not be revealed} to users. App functionality may be at risk as complete control over model outputs may interfere with the content a user may see on the app. For example, Instagram may detect concepts like nudity and violence to ensure that they do not show users any inappropriate content. Giving users access to these concepts could cause irreparable harm to users on the platform. For example, if Instagram or TikTok were to use AI/ML processes on local user data to prevent inappropriate content from being uploaded or age verification it would not be beneficial to allow for user control. Future work needs to determine how to discern between AI/ML for application safety and recommendations.

\subsection{Limitations}
\label{sec:tech_limits}
Our work is limited by a set of technical limitations. 
Both applications included in this study employ techniques to determine if a mobile device is rooted when signing in.
Currently, there does not exist a way to circumvent these checks. 
Due to applications being able to detect when a device is rooted, we \textbf{could not} inform participants of the purpose of the collected data.
Popular anti-hooking tools like AppDome~\cite{appdomePreventDynamic} can detect and alert the application of modern dynamic analysis techniques.
Google also offers their Play Store Integrity API~\cite{androidPlayIntegrity} to validate if a device is authentic.
Rooted devices are not considered authentic and there does not exist a way to bypass this check~\cite{githubGitHubChiteromanPlayIntegrityFix}.
Both applications use the Play Integrity API and AppDome-like dynamic analysis prevention tools.
The AI/ML tools we explored in this work \textit{were not} protected using these tools, allowing us to build interfaces with them for the participants.
However, functions that do utilize these methods downloaded the models, and protected vital components of other areas within the application, limiting our reverse engineering analysis.
We acknowledge this limitation and leave this problem for future work as research surrounding Android improves.

A set of factors, other than technical challenges, limits our work. First, we recruited the participants through our institution's large-scale email service, which might have created a regional bias.
We also had participants self-report, for example, their own AI/ML familiarity and behavioral changes, which could potentially cause a bias.
Therefore, our research does not generalize to all users of social media.
Our study population has a much higher representation of educated individuals.
We acknowledge that our final sample of participants' has a higher education level overall compared to the educational distribution in social media users in general~\cite{hootsuite2024Instagram,statistaUSSocial}.
However, our sample is consistent with other studies performed in a university setting~\cite{andalibi2020human,eslami2016first}.
Nevertheless, future work is needed to, for example, further investigate how people with less exposure to technology demand transparency and statistically quantify the differences in different populations, based on our observations that they were unaware of the detailed AI model implementations, e.g., how the output is generated.

Another limitation is the specificity of the machine learning models provided in our study. Since we are using real machine learning models deployed by social media applications, we are limited to those models.
We only have an opaque understanding of the models themselves and do not know where the data goes after the model gives an output.

\section{Conclusion}

In conclusion, we explored user perceptions and behaviors before and after revealing genuine AI/ML used by Instagram and TikTok.
We learned our participants' reactions depended on how much the AI/ML model affected them once learning about it.
Participants who posted frequently and/or who interacted with the camera were more likely to react negatively.
We also learned that participants felt that \textit{user interactions} were the main source of input for social media AI/ML algorithms.
This influenced participant reactions as they were surprised that the camera could analyze them without their knowledge.
They were also surprised by the \textit{depth} of the model's output being observed with Instagram's model which lead to several assumptions about the purpose of the data.
Our study reported the short-term and long-term behaviors from the participants.
We found that short-term behaviors were likely rooted in participant \textit{usage habits}.
As for long-term behaviors, we found that 8 out of 21 participants indicated they implemented some long-term behavior change after the study.
Our study indicates that the current state of social media algorithm transparency was the main contributor to the confusion harbored by participants.
The lack of transparency lead to participants feeling like the AI/ML models were non-consensual, confusing, creepy, and overwhelming.
So much so that eight individuals spent less time on the apps after being exposed to the models.

\bibliographystyle{ACM-Reference-Format}
\bibliography{refs}

\clearpage

\section{Appendix}

\begin{table}[htbp]

  \Description{Table 2 represents our codebook. There are 5 columns: Themes, Codes, Definition, Example Sub-Codes. }
\centering
\resizebox{\textwidth}{!}{%
  \begin{tabular}{p{0.1\textwidth} p{0.1\textwidth} p{0.2\textwidth} p{0.25\textwidth} p{0.3\textwidth}}
  \toprule
  \textbf{Theme} & \textbf{Sub-theme} & \textbf{Codes} & \textbf{Definition} & \textbf{Example Sub-Codes} \\
  \midrule
  Assumption & ML & Adoption & Participants' adoption of ML models in their lives. & Use for business; Used a lot \\
  &  & Capabilities & Participants' original assumption of ML model capabilities. & Data analysis; Provide advertisements \\
  &  & Concern & Participants' concern towards ML models at the beginning of the interview. & Data analysis; Provide advertisements \\
  &  & Non-Concern & Participants were not concerned about ML models at the beginning of the interview. & Science task; Likes recommendations \\
  &  & Confusion & Participants were confused about what ML is or how ML works. & Unsure how ML is used for advertising; Confused about what pictures are real or fake\\
  &  & Exposure & Ways of the participants learning about ML models. & Work; None \\
  &  & Data Collection & The kind of data participants are considering to be collected by ML models. & Demographics; Interests \\
  & Instagram & Model & Participants' assumption of IG model's usage. & Human classification; Location estimation \\
  &  & Data Usage & Participant's assumption of how IG uses the information collected from the model. & Saved to database; Auto tag images\\
  &  & Data Collection & The kind of data participants are assuming to be collected by the IG model. & Ask for location when editing photos; People take photos outside\\
  &  & Concern & Concerns participants hold towards IG's data collection behavior they assumed. & Too much data in IG already; Want to be private\\
    & TikTok & Algorithm & Participants' assumption towards TikTok's algorithm. & Feel to personal; Live updates to recommend better content \\
  &  & Concern & Participants' concern towards the TikTok model they assumed. & Low accuracy; Would not want anyone estimating gender \\
  &  & Data Collection & Participants' assumption of how TikTok uses the data being collected. & Sold to advertisers; App development \\
  &  & Model & Participants' assumption on TikTok's ML model. & Accuracy varies; Race can also be assumed with facial features \\
  &  & Capabilities & Participants assumption on TikTok's ML model's capability. & Feasible; Does not believe \\

  \bottomrule
  \end{tabular}
}
\caption{Themes.}
\label{tab:theme1}
\end{table}

\clearpage

\begin{table}[htbp]
\Description{Table 3 represents our codebook. There are 5 columns: Themes, Codes, Definition, Example Sub-Codes. }
\centering
\resizebox{\textwidth}{!}{%
  \begin{tabular}{p{0.1\textwidth} p{0.1\textwidth} p{0.2\textwidth} p{0.25\textwidth} p{0.3\textwidth}}
  \toprule
  \textbf{Theme} & \textbf{Sub-theme} & \textbf{Codes} & \textbf{Definition} & \textbf{Example Sub-Codes} \\
  \midrule
  & Social Media & Advertising & Participants' assumption on social media would advertise using collected data. & People get targeted ads; Advertising increased in TikTok\\
  &  & Algorithm & Participants' assumption on social media's algorithm. & Recommendations come from past usage; Only use data if accuracy is over 99\%\\
  &  & Behavior & Participants' assumed behavior change on their social media usage. & Will be more aware; No change because of rarely posting\\
  &  & Reflection & Participants' reflection on social media using ML models. & Opinion hinges on data usage; We are forced to use tech as a society\\
  &  & Concern & Participants' concerns towards social media's ML usage they assumed. & Things appearing to be real; Nefarious data use\\
  &  & Non-Concern & Participants are not concerned about social media's ML usage they assumed. & Usage Statistics; App improvements only\\
  &  & Data Collection & Participants assumed items that social media collects. & Actions humans do; Knows faces well\\
  &  & Data Usage & Participants assumed how social media would use the information they collected. & Necessary for profit; Know who is using the app\\
  Reaction & ML & Lesson Reaction & Participants' reaction towards the ML lesson we presented during the interview. & Interesting that it can detect no person; Photos are gathered so quickly \\
  & Instagram & Concern & Participants' concerns towards the IG model we presented. & Granularity is concerning; Model does not meet expectations \\
  &  & Behavior & Participants' behavior changed after seeing the IG model we presented. & Will still use the app; Will be more aware \\
  &  & Model & Participants' reaction towards the IG model we presented. & High accuracy; High confidence values seem correct\\
  &  & Non-Concern & Participants did not express concern towards the IG model we presented. & Comfortable with model; Nothing to hide not concerned\\
  &  & Reflection & Participants' reflection after seeing the IG model we presented. & Opinions could change; Sobering due to amount of data\\
  &  & Labels & Participants' reaction on the labels shown on IG model. & So many labels; Seems focused on western ideals\\
  & TikTok & Accuracy & Participants' reaction on how they rate the accuracy of the TikTok model we show. & Terrifying; Gender estimation is correct \\
  &  & Concern & Participants' concern towards the TikTok model we show. & Age and gender estimation is not generalizable; Different demographics would react differently\\
  &  & Expectations & Participants' expectations towards TikTok's model. & Lives up to expectations; Technically yes but privacy no\\
  &  & Label & Participants' reaction towards the label on TikTok's model. & Shocked with age; Gender identification is subjective\\
  &  & Model & Participants' immediate reaction toward the TikTok model we show. & A human would do better; Has a worse opinion of TikTok\\
  &  & Behavior & Participants' behavior change after seeing TikTok's model we show. & Will not re-download the app; Temporarily change behavior\\
  &  & Reflection & Participants' reflection on the TikTok model we show. & Positive if they give option; Data collection is similar to colonialism's racism\\

  \bottomrule
  \end{tabular}
}
\caption{Themes (continue).}
\label{tab:theme2}
\end{table}

\clearpage

\begin{table}[htbp]
\Description{Table 4 represents our codebook. There are 5 columns: Themes, Codes, Definition, Example Sub-Codes. }
\centering
\resizebox{\textwidth}{!}{%
  \begin{tabular}{p{0.1\textwidth} p{0.1\textwidth} p{0.2\textwidth} p{0.25\textwidth} p{0.3\textwidth}}
  \toprule
  \textbf{Theme} & \textbf{Sub-theme} & \textbf{Codes} & \textbf{Definition} & \textbf{Example Sub-Codes} \\
  \midrule
  Data Collection & Instagram & Concern & Participant's concern towards IG's data collection. & Bad accuracy leads to bad recommendations; Concerned with clothing detection\\
  &  & Confusion & Participant's confusion on IG's data collection. & Data collection does not seem  beneficial; Unaware of uses\\
  &  & Expectation & Participant's expectation towards IG's data collection. & Data collected would only benefit business\\
  \addlinespace
  Usage & Social Media & Assumption & Participants' assumption towards social media usage. & Best place for portfolio; Does not see posts from friends and family\\
  &  & Content Engagement & Participants' content engagement when they use social media. & Does not live stream; Does not use camera\\
  &  & Posting Habits & Participant's posting habits on social media. & Rarely posts; Posting less when older\\
  &  & Social Connections & Participants use social media for social connections. & Maintains connections not build an audience; Friends\\
  Privacy and Transparency & Instagram & Concern & Participants' privacy concern towards IG. & Famous people would care more about privacy; Would feel like I was being watched \\
  &  & Perception & Participants' perception of privacy in IG. & Depends on circumstances; Depends on the person \\
  &  & Non-Concern & Participants do not have privacy concerns about IG. & Privacy overall is not a concern; Not worried about labels\\
  &  & Terms of Service & Participants' understanding of Terms of Service in IG. & Wants to be notified of this in terms; Thinks they should be more specific\\
  &  & Awareness & Participants call for awareness of privacy issues. & People should be aware; Does want to know\\
  & TikTok & Terms of Service & Participants' understanding of Terms of Service in TikTok. & They have to let us know; If its fine print then its user's fault\\
  &  & Transparency & Participants are not satisfied with the transparency in TikTok's privacy settings. & Children may not be able to consent; Should be transparent\\
  &  & Concern & Participants' privacy concern towards TikTok. & I'll be a target; Assumed camera frames were not analyzed\\
  & Social Media & Assumption & Participants' assumption on social media privacy terms. & More aware if told; Everyone would disable ML \\
  &  & Limitations & Participants consider social media's functionality to be limited. & Doesn't know if being able to toggle is possible; Algorithm quality could drop \\
  &  & Non-Concern & Participants do not have privacy concerns about social media. & Purchased items; Consumption patterns \\
  &  & Concern & Participants' privacy concern towards social media. & Info can be gleaned from image background; Gore or violence\\

  \bottomrule
  \end{tabular}
}
\caption{Themes (continue).}
\label{tab:theme3}
\end{table}

\clearpage

\begin{table}[htbp]
\Description{Table 5 represents our codebook. There are 5 columns: Themes, Codes, Definition, Example Sub-Codes. }
\centering
\resizebox{\textwidth}{!}{%
  \begin{tabular}{p{0.1\textwidth} p{0.1\textwidth} p{0.2\textwidth} p{0.25\textwidth} p{0.3\textwidth}}
  \toprule
  \textbf{Theme} & \textbf{Sub-theme} & \textbf{Codes} & \textbf{Definition} & \textbf{Example Sub-Codes} \\
  \midrule
  &  & Responsibility & Participants consider how to distribute the responsibility of keeping privacy in social media. & Public posts are ok; Do not post to avoid ML\\
  &  & Terms of Service & Participants' perception of Terms of Service of social media. & SM companies require our data for profit, which is in the terms of service; No one reads them\\
  \addlinespace
  Deployment & Social Media & Assumption & Participants' assumption on what functionality social media is deploying ML model on.  & More useful for them after posting; Only use ML with face filters\\
  & & Capabilities & Participants' guessing on what functionality social media is capable of deploying ML model on. & Confident in usage for face filters; Confident with editing photo and face filters\\
  & & Confusion & Participants' confusion on how social media could deploy the model in specific functionality. & Too complicated to comprehend; Unsure about opening photo\\
  & & Concern & Participants' concern towards these ML model deployments. & Not comfortable with using after posting; Should be better before deploying\\
  & & Non-Concern & Participants do not hold concerns towards the ML model deployments. & Fine with all instances; Learning from photo editing is fine\\
  & & Expectations & Participants' expectations on what functionality is acceptable to deploy ML models on. & My device = no data for them; Face filter, editing photo, and after posting\\
  \addlinespace
  Comparison & Social Media & Concern & Participants comparing the concerns they hold for IG and TikTok. & Gender and age feel unsettling; TikTok feels less consensual than IG\\
  &  & Non-Concern & Participants are comparing and are not concerned about IG or TikTok. & Already assumed this was happening; Data collected is not a problem at all\\
  &  & Model & Participants are comparing the ML models on IG and TikTok. & Instagram looks for background information; TikTok looks for less labels\\
  &  & Preference & Participants' preference comparing IG and TikTok ML models. & Instagram; TikTok\\
  &  & Reflection & Participants' reflection comparing IG and TikTok ML models. & Seeing the models grounded their perspective; Reaction to models not being public\\
  
  \bottomrule
  \end{tabular}
}
\caption{Themes (continue).}
\label{tab:chatgpt_risks}
\end{table}

\end{document}